\documentclass[11pt]{article}

\usepackage{amsmath}
\usepackage{amssymb}
\usepackage{graphicx}

\usepackage{color} 

\usepackage[labelfont=bf,labelsep=period,justification=raggedright]{caption}

\usepackage{soul}
\usepackage[utf8]{inputenc}

\usepackage[comma, round,sort]{natbib} 
\date{}

\newcommand{\F}{\mathcal{F}}
\newcommand{\Prob}{\mathbb{P}}

\title{Using haplotype differentiation
      among hierarchically structured populations for the detection of
      selection signatures}

\author{%
Fariello Mar\'ia In\'es$^{\ast,\S,\ast \ast}$, 
Boitard Simon$^{\ast}$,
Naya Hugo$^{\S,\S \S}$,
SanCristobal Magali$^{\ast}$, 
Servin Bertrand$^{\ast}$\\
}
\begin{document}
\maketitle
\begin{flushleft}
  {$\ast$} Laboratoire de Génétique Cellulaire, INRA, Toulouse, France\\
  {$\S$} Unidad de Bioinform\'atica, Institut Pasteur, Montevideo, Uruguay\\
  {$\ast \ast$} Facultad de Ingenier\'ia, Universidad de la República, Montevideo, Uruguay\\
  {$\S$} Facultad de Agronom\'ia, Universidad de la República, Montevideo, Uruguay \\
\end{flushleft}

\newpage

\begin{flushleft} 

{{\bf Running title:}} Haplotype-based detection of selection signatures \\

{{\bf Key-words:}}  selective sweeps; haplotype; linkage disequilibrium; $F_{ST}$; genome scan; sheep;
statistical genetics; structured populations; population genetics

{{\bf Corresponding Author:}}\\ Mar\'ia In\'es Fariello,\\
Auzeville, BP 52627  Chemin de Borde Rouge\\
31326 Castanet Tolosan Cedex,\\
France\\
05 61 28 54 35\\
E-mail: mfariello@toulouse.inra.fr

\end{flushleft}

\newpage

\section*{ABSTRACT}

The detection of molecular signatures of selection is one of the major
concerns of modern population genetics. A widely used strategy in this
context is to compare samples from several populations, and to look
for genomic regions with outstanding genetic differentiation between
these populations. Genetic differentiation is generally based on
allele frequency differences between populations, which are measured
by $F_{ST}$ or related statistics. Here we introduce a new statistic,
denoted $hapFLK$, which focuses instead on the differences of
haplotype frequencies between populations. In contrast to most
existing statistics, $hapFLK$ accounts for the hierarchical structure
of the sampled populations.  Using computer simulations, we show that
each of these two features - the use of haplotype information and of
the hierarchical structure of populations - significantly improves the
detection power of selected loci, and that combining them in the
$hapFLK$ statistic provides even greater power. We also show that
$hapFLK$ is robust with respect to bottlenecks and migration and
improves over existing approaches in many situations.  Finally, we
apply $hapFLK$ to a set of six sheep breeds from Northern Europe, and
identify seven regions under selection, which include already reported
regions but also several new ones. We propose a method to help
identifying the population(s) under selection in a detected region,
which reveals that in many of these regions selection most likely
occurred in more than one population. Furthermore, several of the
detected regions correspond to incomplete sweeps, where the favorable
haplotype is only at intermediate frequency in the population(s) under
selection.

\section*{INTRODUCTION}

The detection of molecular signatures of selection is one of the major
concerns of modern population genetics. It provides insight on the
mechanisms leading to population divergence and differentiation. It
has become crucial in biomedical sciences, where it can help to
identify genes related to disease resistance \citep{FUM2010,ALB2010,
  CAG2008, NatSel, TIS2001}, adaptation to climate
\citep{REE2012,STU2009, LAO2007} or altitude \citep{BIG2010,
  SIM2010}. In livestock species, where artificial selection has been
carried out by men since domestication, it contributes to map traits
of agronomical interest, for instance related to milk \citep{HAY2009}
or meat \citep{kijas-etal-12} production.

Efficiency of methods for detecting selection vary with the considered
selection time scale \citep{Sabeti}. For the detection of selection
within species (the ecological scale of time), methods can be
classified into three groups : methods based on (i) the high frequency
of derived alleles and other consequences of hitchhiking within
population \citep{BOI2009, KIMSTE2002, KIMNIE2004, NIE2005}, (ii) the
length and structure of haplotypes, measured by EHH or EHH derived
statistics \citep{EHH, iHS} and (iii) the genetic differentiation
between populations, measured by $F_{ST}$ or related statistics
\citep{LK73,Maxime,BeaumontBalding,FollGag,Riebler,Gautier}. Methods
of the latter kind, which we will be focusing on, are particularly
suited to the study of species that are structured in well defined
populations, such as most domesticated species. In contrast to methods
based on the frequency spectrum (i) or the excess of long haplotypes
(ii), they can detect a wider range of selection scenarios, including
selection on standing variation or incomplete sweep, albeit up to a
given extent \citep{YI2010, INN2008}.

The most widely used statistic to detect loci with outstanding genetic
differentiation between populations is the $F_{ST}$ statistic
\citep{FstH, NatSel}.  The general application of $F_{ST}$-based scan
for selection is to identify outliers in the empirical distribution of
the statistics computed genome-wide. One major concern with this
approach is that it implicitly assumes that populations have the same
effective size, and derived independently 
from the same ancestral population, \emph{i.e.} with a star-like
evolution tree. If this hypothesis does not hold, which is often the
case, genome scans based on raw $F_{ST}$ can suffer from bias and
false positives, an effect that is similar to the well known effects
of cryptic structure in genome-wide association studies
\citep{PRI2010}. To cope with this problem several methods have been
proposed to account for unequal population sizes
\citep{BeaumontBalding,FollGag,Riebler,Gautier}, however few solutions
have been proposed to deal with the hierarchical structure of
populations \citep{Excoffier}. Among them \citet{Maxime} proposed an
extension of the classical Lewontin and Krakauer (LK) test
\citep{LK73}, where the hierarchical population structure is
captured through a kinship matrix, which is used to model the
covariance matrix of the population allele frequencies. A similar
covariance matrix was also introduced in a related context to account
for the correlation structure arising from population geography
\citep{Coop}.

All $F_{ST}$ based approaches discussed above are single marker tests,
\emph{i.e.} markers are analyzed independently from each other.  As
dense genotyping data and sequencing data are now common in population
genetics, accounting for correlations between adjacent markers has
become necessary. Furthermore, haplotype structure contains useful
information for the detection of selected loci, as demonstrated by the
within-population methods mentioned above (class (ii)). Several
strategies for combining the use of multiple populations and of
haplotype information have thus been proposed recently. These include
the development of EHH related statistics for the comparison of pairs
of populations \citep{XP-EHH,Rsbi}, the introduction of dependence
between SNPs (Single Nucleotide Polymorphisms) in $F_{ST}$-based
approaches through autoregressive processes \citep{Gompert, Guo}, or
the computation of $F_{ST}$ using local haplotype clusters that are
considered as alleles \citep{Browning}. However, none of these
approaches accounts for the possibility that populations are
hierarchically structured.

We present here an haplotype-based method for the detection of
positive selection from multiple population data. This new statistic,
$hapFLK$, builds upon the original $FLK$ statistic \citep{Maxime}. As
$FLK$, it incorporates hierarchical structure of populations, but the
test is extended to account for the haplotype structure in the
sample. For this, it uses a multipoint linkage disequilibrium model
\citep{FPh} that regroups individual chromosomes into local haplotype
clusters. The principle is to exploit this clustering model to compute
``haplotype frequencies'' which are then used to measure
differentiation between populations. The idea of using localized
haplotype clusters to study genetic data on multiple populations has
been proposed before \citep{Browning, JAC2008}.  \citet{Browning}
showed that using haplotype clusters rather than SNPs allowed to
circumvent, to some extent, the problems arising from SNP
ascertainment bias. They also showed that two genome regions known to
have been under strong positive selection in particular human
populations exhibited large population specific haplotype-based
$F_{ST}$. \citet{JAC2008} showed by using fastPHASE that there was a
predominance of a single cluster haplotype in the Hapmap population of
Utah residents with ancestry from northern and western Europe (CEU
population) in the region of the LCT gene and interpreted this signal
as a recent selective sweep.

In this paper, we examined in detail the ability of statistics based
on population differentiation at the haplotype level to capture
selection signals.  Using computer simulations, we study the power and
robustness of our new haplotype based method for different selection
and sampling scenarios, and compare it to single marker ($F_{ST}$ and
$FLK$ \citep{Maxime}) and haplotype based (XP-EHH \citep{XP-EHH})
approaches.  To illustrate the interest of this approach, we provide a
practical example on a set of 6 sheep breeds for which dense
genotyping data has been recently released by the Sheep HapMap project
\citep{kijas-etal-12}. 
In this context, we propose a new strategy for the detection of
outliers loci in genome scans for selection and describe a method for
the identification of the populations that have experienced selection
at a detected region.

\section*{METHODS}

\subsection*{Test Statistics}

\paragraph{$\mathbf{F_{ST}}$ and $\mathbf{FLK}$ tests for SNPs:}
Consider a set of $n$ populations that evolved without migration from
an ancestral population, and a set of $L$ SNPs in these
populations. For a given SNP, let $p=(p_1,\ldots, p_i, \ldots , p_n)'$
be the vector of the reference allele frequency in all
populations. Denoting $\overline{p}$ and $s^2_p$ the sample estimates
of the mean and variance of the $p_i$'s, the Fisher's fixation index
$F_{ST}$ at this SNP is given by
$\frac{s^2_p}{\overline{p}(1-\overline{p})}$. This index quantifies
the genetic differentiation between populations and is commonly used
to detect loci under selection. Loci with outstanding high (resp. low)
values of $F_{ST}$ can be declared as targets of positive
(resp. balancing) selection.

However, if the sampled populations have unequal effective sizes
or/and are hierarchically structured, genome scans based on raw
$F_{ST}$ values can bias inference.  For instance, a given allele
frequency difference between two closely related populations should
provide more evidence for selection than the same difference between
two distantly related populations. To account for these drift and
covariance effects when detecting loci under selection, \citet{Maxime}
introduced the statistic
\begin{equation} \label{FLK}
T_{FLK} = (p - p_0 \textbf{1}_n )' Var(p)^{-1}(p - p_0 \textbf{1}_n )
\end{equation}
where $p_0$ is the allele frequency in the ancestral population and
$Var(p)$ is the expected covariance matrix of vector $p$, which they
modeled as:
\begin{equation} \label{F}
Var(p) = \F p_0 (1- p_0)
\end{equation}
$\F_{i,i}$ is the expected inbreeding coefficient in population
$i$ and $\F_{i,j}$ is the expected inbreeding coefficient in the
ancestral population common to populations $i$ and $j$. In other
words, the entries of the kinship matrix $\F$ represent the amount of
drift accumulated on the different branches of the population
tree. They can be derived as a function of the divergence times and
the effective population sizes along the population tree, as described
in Supporting Information (SI) section 1.1.

In practice, these demographic parameters are unknown and $\F$ must be
estimated from genome wide data. Here, it is done as follows: first,
pairwise Reynolds' distances \citep{Reynolds} between populations
(including an outgroup) are computed for each SNP and averaged over
the genome.  Then, a phylogenetic tree is fitted from these distances
using the Neighbor-Joining algorithm.  The branch lenghts of this tree
are finally combined to compute $\F$ entries.  More details on this
procedure can be found in \citet{Maxime}.  Given the estimation of
$\F$, and unbiased estimator of $p_0$ is obtained as:
$$
\hat{p}_0 = \frac{\textbf{1}_n' \F ^{-1}p}{\textbf{1} _n' \F ^{-1} \textbf{1} _n} = w' p
$$
and can be used in equations (\ref{FLK}) and (\ref{F}) to obtain
$T_{FLK}$.

Under the assumption that all populations diverged simultaneously from
the same ancestral population (star like evolution) and with the same
population size, $\F$ is equal to $\overline{F}_{ST} I_n$, where
$\overline{F}_{ST}$ is the average $F_{ST}$ over all SNPs and $I_n$ is
the identity matrix of size $n$.  In this case, $T_{FLK}$ is
equivalent to the LK statistic \citep{LK73}

\[T_{LK} = \frac{n-1}{\overline{F}_{ST}}F_{ST} \]

\paragraph{$\mathbf{FLK}$ test for multiallelic markers:} Considering
haplotypes as multiallelic markers, an extension of the $FLK$
statistic in the case where each locus presents more than 2 alleles is
required. Let $A$ be the number of alleles at a given locus, the
allele frequency vector becomes:
$$
P = ( \underbrace{p_{1,1}, \ldots , p_{1,n}}_{allele~1}, \ldots,\underbrace{p_{A,1}, \ldots, p_{A,n}}_{allele~A} )' 
= (p_{1}, \ldots, p_{A})'
$$
and a multiallelic version of the $T_{FLK}$ statistic is provided by
\begin{equation} \label{FLKmulti}
T_{FLK} = (P - P_0  \otimes \textbf{1}_n )' Var(P)^{-1}(P - P_0 \otimes \textbf{1}_n )
\end{equation}
where $\otimes$ denotes the Kronecker product and $P_0=(p_{1,0},
\ldots, p_{A,0})'$ contains the allele frequencies of the $A$ alleles
in the ancestral population.  $Var(P)$ is written:
\begin{equation} \label{var}
\begin{array}{rll} Var(P) =& \left( \begin{array}{ccc} Var(p_{1}) & \cdots & Cov(p_{1}, p_{A}) \\
			\vdots & Var(p_{a}) & 	\vdots \\
			Cov(p_{A},p_{1}) & \cdots & Var(p_{A})\\		
\end{array} \right) & = \mathcal{B}_0 \otimes \mathcal{F}, \end{array}
\end{equation}
with $\mathcal{B}_0= diag(P_0) - P_0 P_0'$. Each diagonal block of
$Var(p)$ corresponds to the biallelic covariance matrix for one of the
$A$ alleles, while the extra-diagonal blocks arise from the covariance
terms between different alleles. Similar to the biallelic case, $P_0$
is estimated by $\hat{P}_0=(w p_{1}, \ldots, w p_{A})\prime$.  $Var(P)$
is inverted using the Moore-Penrose generalized inverse.

\paragraph{$\mathbf{FLK}$ test for haplotypes:}

The \citet{FPh} model summarizes local haplotype diversity in a sample
through a reduction of dimension by clustering similar haplotypes
together. These clusters can then be considered as alleles to compute
the haplotype version of $T_{FLK}$ statistic. Let $g_i^{\ell}$ be the
genotype observed for individual $i$ at marker $\ell$. In the Hidden
Markov Model of \citet{FPh}, $g_i^{\ell}$ is associated to a hidden
state $z_i^{\ell}=(z_i^{\ell,1},z_i^{\ell,2})$, where $z_i^{\ell,1}$
and $z_i^{\ell,2}$ represents the pair of clusters giving rise to the
(diploid) individual genotype. The Markov structure of $z_i=(z_i^1,
\ldots, z_i^L)$ along the genome implies that cluster memberships of
close markers are correlated, which allows to account for linkage
disequilibrium effects. When this model is fitted to the whole
genotype data $g$, it provides for each individual $i$, marker $\ell$
and cluster $k$, the posterior probabilities $\Prob (z^{\ell,1}_{i} =
k \vert g,\Theta )$ and $\Prob (z^{\ell,2}_{i} = k \vert g,\Theta )$,
where $\Theta$ is a vector of estimated model parameters (see
\citet{FPh} for more details). Cluster probabilities in each
population $j$ are obtained by averaging the probabilities of the
$n_j$ individuals of this population, {\it i.e.}:
\begin{equation}
p^{\ell}_{k,j}= \frac{1}{2 n_j} \sum_{i = 1}^{ n_j} (\Prob (z^{\ell,1}_{i} =k \vert g,\Theta ) + \Prob (z^{\ell,2}_{i} =k \vert g,\Theta ) )
\end{equation}
Considering clusters as alleles and population-averaged probabilities
as population frequencies, the allele frequency vector of a marker
$\ell$ is:
\[ P^l=\left(\underbrace{p_{1,1}^{\ell}, \dots , p_{1,n}^{\ell}}_{cluster~1}, \underbrace{p_{2,1}^{\ell}, \dots,p_{2,n}^{\ell}}_{cluster~2}, \dots, \underbrace{
p_{K,1}^{\ell} ,\dots, p_{K,n}^{\ell} }_{cluster~K}\right)' \]

For each marker $\ell$, the multiallelic statistic $T_{FLK}$ is
computed according to equation (\ref{FLKmulti}), with a small
modification in the derivation of $Var(P)$.  Indeed, clusters that are
fitted in the present population can not exactly be considered as real
alleles that already existed in the ancestral population, as assumed
by the original $T_{FLK}$ statistic.  Moreover, the generalized
inverse of $\mathcal{B}_0 \otimes \mathcal{F}$ was found numerically
unstable for small $p_{a,0}$ values, which are very common when the
number of alleles is large.  Consequently, the $\mathcal{B}_0$ matrix
is replaced by the identity matrix $I_{A}$ in equation (\ref{var}),
leading to the statistic
\begin{equation} \label{hapFLKB}
T_{FLK} = (P - P_0  \otimes \textbf{1}_n )' (\mathcal{I} \otimes \mathcal{F} )^{-1}(P - P_0 \otimes \textbf{1}_n )
\end{equation}
Simulations confirmed that this version of the test was more powerful
than the one including $\mathcal{B}_0$ (SI, Figure S1).

For the model of \citet{FPh}, parameter estimates and cluster
membership probabilities are obtained using an expectation
maximization (EM) algorithm.  Because this algorithm converges to a
local maximum, it is recommended to run it several times from
different starting points. Applying the model to haplotype phasing,
\citet{FPh,guan.stephens.2008} observed that averaging the results
from these different runs was more efficient than keeping the maximum
likelihood run, which may be due to the fact that different runs are
optimal in different genomic regions.  Following their strategy, we
averaged the statistics obtained using equation (\ref{hapFLKB}) from
different EM iterations to finally obtain the haplotype extension of
$FLK$.  We denote this extension $hapFLK$.

The haplotype extension of the $F_{ST}$ test, denoted $hapF_{ST}$ in the
simulation study, was obtained by replacing $\mathcal{F}$ by $I_n$ in
equation (\ref{hapFLKB}), therefore ignoring the hierarchical
structure of populations.

Software implementing the $hapFLK$ calculations is available at 

\texttt{https://forge-dga.jouy.inra.fr/project/hapflk }

\subsection*{Simulations}

To evaluate the performance of $hapFLK$ and compare it to that of
other tests, we performed a set of simulations mimicking the data
obtained from dense SNP genotyping or full sequencing of samples from
multiple populations. In particular, we designed our simulation to
match the data produced within the Sheep HapMap project
\citep{kijas-etal-12} (analyzed below), in terms of population
divergence and SNP density.

\begin{figure}[htp]
\begin{center}
\includegraphics[width=3.3in]{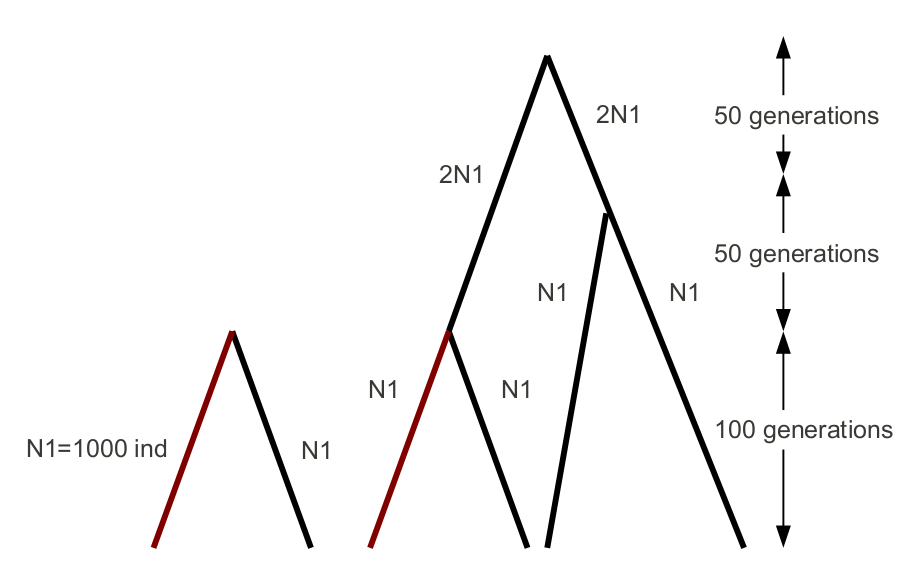}
\end{center}
\caption{ { \bf Population trees for the two simulated scenarios.} The red
  branch indicates the selected population and time during which
  selection acts.}
\label{Figure1}
\end{figure}

\paragraph{Scenarios with constant size and no migration:}
Two scenarios were simulated one with 2 populations and the other one
with 4 populations (Figure \ref{Figure1}).  The 2-population scenario
was designed to be a subtree of the 4-population scenario, which
allows to compare the detection power obtained by testing the 4
populations jointly, with that obtained by testing all possible pairs
of populations.

The ancestral population was simulated using using {\it ms}
\citep{Hudson}, with mutation rate $ \mu= 10^{-8}$, recombination rate
$c=10^{-8}$ (1 centiMorgan per megabase), and region length $L= 5
Mb$. The effective population size and the number of simulated
haplotypes were $N_e=1000$ and $n_h = 4000$ for the 2 population case,
and $N_e=2000$ and $n_h = 8000$ for the 4 population case. The
generated haplotypes had around 200 SNPs per Mb. The first two
populations (top branches in Figure \ref{Figure1}) were created
independently by sampling half of the individuals from the founder
population. A forward evolution of the populations after their initial
divergence was then simulated with the \emph{simuPOP} Python library
\citep{simuPOP}, under the Wright-Fisher model. During forward
simulations, recombination was allowed but mutation was not.


Simulations were performed with and without selection. For scenarios
with selection, selection occurred at a single locus, in the red
branch shown in Figure \ref{Figure1}. The selected locus was chosen as
the closest to the center of the simulated region, among the SNPs with
minor allele frequency equal to a predefined value ($ 0.01, 0.05, 0.10, 0.20
$ or $ 0.30 $). The less frequent allele of this SNP was given fitness
$1+s$, with selection intensity $s=0.05$ (leading to $\alpha = 2 \cdot
N_e \cdot s = 100 $). Individual's fitness was $1$ for homozygotes
with the non selected allele, $1+s$ for heterozygotes and $(1+s)^2$
for homozygotes with the selected allele.  

At the end of each simulation replicate 50 individuals were sampled
from each of the final populations, and SNPs with a Minor Allele
Frequency (MAF) greater than $5\% $ kept.  Two different genotyping
densities were considered: 20 SNPs per Mb (equivalent to that of 60K
SNPs in sheep) and 100-125 SNPs per Mb (all remaining SNPs).  The
statistics $T_{F_{ST}}, T_{hapF_{ST}}, T_{FLK}$ and $T_{hapFLK}$ were
computed at each SNP, assuming that the kinship matrix $\F$ was known.
Indeed, the estimation of $\F$ is very accurate for evolution
scenarios with constant population size and no migration (see
\citet{Maxime} and SI Figure S2).  Parameters used
for running the test were $K=5$ (number of clusters) and $em=5$
(number of EM runs) for the two-population scenario and $K=20$ and
$em=5$ for the four-population scenario.  These values were chosen for
maximizing the detection power. Greater values did not improve this
power, but increased computation time.  For the two-population
scenario, the $XP-EHH$ statistic \citep{XP-EHH} was also computed at
each SNP, using software obtained from
\texttt{http://hgdp.uchicago.edu/Software/}.

Power of the tests was computed as follows.  3000 data sets were
simulated under the null (neutrality) and 3000 under the alternative
(selection) hypotheses, for each scenario considered. In simulations
under selection, only replicates where the final frequency of the
selected allele was greater than $ 60 \%$ were kept.  For each
replicate and statistic $S$,the maximum value $S^{max}$ over the $5
Mb$ region was recorded.  This provides the distribution of $S^{max}$
under the null and the alternative hypotheses. The power of a test
with statistic $S$, for a given type I error $\alpha$ is the
proportion of simulations under selection for which $S^{max} >
q_{\alpha}$ where $q_{\alpha}$ is the (1-$\alpha$)th quantile of the
null distribution of $S^{max}$.


\paragraph{Scenarios with bottlenecks or migrations:}

To study the robustness of the approach, more complex demographic
events were investigated through three scenarios.  They derived from
the 2 population scenario described above, with the following
modifications: (i) A bottleneck in a single population: the effective
size in this population was set to $N_e=100$ in the first 5
generations following the split, and to $N_e=1852$ in later
generations; (ii) Asymmetric migration: at generation 51, population 1
sent 10 \% of migrants to population 2; (iii) Symmetric migration: at
generation 51, population 1 sent 10 \% of migrants to population 2 and
recieved 10\% of migrants from population 2.  In terms of expected
drift at a single SNP, these scenario are equivalent to the constant
size scenario (see SI section 1.1 for a proof). Hence,
they can be used to evaluate the influence of the underlying
demographic model on $hapFLK$, while conditioning on a fixed value of
$\F$.  To ensure that the $\F$ matrix used in $hapFLK$ fits the one
that would be estimated from real data, 100 artificial whole genome
dataset were created for each of the scenarios (i)-(iii), and used to
estimate $\F$.  Each artificial whole genome datasets was created by
simulating 500 independent genome segments of 5 Megabases.

Robustness of $hapFLK$ and XP-EHH were evaluated by comparing
quantiles of each statistic obtained under bottleneck or migration
demography with those obtained under a constant size evolution. No
selection was applied in these simulations.

Evaluation of the detection power of $hapFLK$ and XP-EHH under
bottleneck (or migration) with selection, was performed as described
above, \emph{i.e.}  distributions obtained under neutrality provided
quantiles used to calibrate type I error. Because scenarios (i) and
(ii) are asymmetric, each one provided two different simulation
scenarios under selection, one with selection in population 1 and one
with selection in population 2.

\subsection*{Sheep data analysis}

A whole genome scan for selection in Sheep was performed using the
genotype data from the Sheep HapMap project (available at \texttt{
  http://sheephapmap.org/download.php}). The Sheep HapMap dataset
includes 2819 animals from 74 breeds, collected in such a way that it
represents most of the worldwide genetic diversity in the
Sheep. Genotypes at 48703 autosomal SNPs (after quality filtering) are
available for these animals. Focus was placed on the North-European
group, all populations with less than 20 individuals being
removed. Populations resulting from a recent admixture were also
excluded because they are not compatible with the population tree
model assumed for our test.  Finally, the following populations were
included in the analysis (sample size in parentheses): Galway (49),
Scottish Texel (80), New Zealand Texel (24), German Texel (46), Irish
Suffolk (55) and New Zealand Romney (24). The Soay breed was used as
an outgroup for computing the $\F$ matrix.

\paragraph{Parameters of the $hapFLK$ analysis:} To determine the
number of clusters to be used in the fastphase model, the
cross-validation procedure of fastPHASE was used which indicated an
optimal number of 45 clusters. As the computational cost increases
quadratically with the number of clusters, and as the genome scans
performed on one single chromosome for 40 and 45 clusters provided
very similar results, 40 clusters were used for the rest of the
analysis. A sensitivity analysis indicated that on this dataset 45 EM
runs were required to get a stable estimate of $hapFLK$.

\paragraph{Computation of p-values:} In contrast to the simulated
datasets, real data does not provide null distribution allowing to
compute p-values from the $hapFLK$ statistics. Also, due to
ascertainment bias in the SNP panel, we believe that performing
neutral simulations based on an estimation of $\F$ is not a good
strategy for this particular dataset (see the Discussion for more
details). P-values were thus estimated using an empirical approach
(described below) exploiting the fact that selected regions, at least
those that can be captured with $hapFLK$, affect a small portion of
the genome.

The genome-wide distribution of $hapFLK$ appeared to be bi-modal, with
a large proportion of values showing a good fit to a normal
distribution, and a small proportion of extremely high values ( SI
Figure S3 ). Consequently, p-values were estimated as follows: first,
robust estimators of the mean and variance of $hapFLK$ were obtained,
to reduce the influence of outliers. For this estimation the
\texttt{rlm} function of the package \texttt{MASS} \citep{VEN2002} in
\emph{R} was used.  $hapFLK$ values were then standardized using these
estimates and corresponding p-values computed from a standard normal
distribution. The resulting distribution of p-values across the genome
was found to be close to uniform for large p-values, consistent with a
good fit to the normal distribution apart from the outliers which
exhibit small p-values.  Using the approach of \citet{STO2003}, the
FDR estimated when calling significant hypotheses with $p<10^{-3}$ was
5\%.


\paragraph{Pinpointing the selected population:} Similar to all
$F_{ST}$ related tests, $hapFLK$ detects genomic regions where genetic
data is globally not consistent with a neutral evolution, but does not
directly indicate where selection occured in the population tree. To
investigate this question, branch lengths of the population tree were
re-estimated for each significant region, using SNPs exceeding the
significance threshold. The principle was to fit (using ordinary least
squares) the branch lengths to the local values of Reynolds genetic
distances. For each branch the p-value for the null hypothesis of no
difference between the lengths estimated from data in the region and
in the whole genome was computed. Details on the procedure are
provided in SI section 1.3.

\section*{RESULTS}

\subsection*{Simulation results}
We performed a set of simulations to evaluate the performance of
$hapFLK$, in comparison with that of other tests (see Methods for more
details).  To present the results of these simulations, we begin with
scenarios that fit the assumptions of our model: a population
tree without migration and with constant size within each branch.  We
then move to more complex demographic scenarios, which are expected to
be less favorable to our test.

\paragraph{Interest of using haplotypes over SNPs:}
We first performed simulations assuming data from two populations of
the same effective size (Figure \ref{Figure1}, left).  In this setting,
the structure-aware tests ($FLK$ and $hapFLK$) are equivalent to their
unaware counterparts ($F_{ST}$ and $hapF_{ST}$ resp.).

In simulations mimicking dense genotyping data, the use of haplotype
information ($hapFLK$) provides more detection power than the use of
single SNP tests ($F_{ST}$). This holds for both hard sweeps ($p_0=0.01$)
and soft sweeps detection ($p_0$ up to $0.3$, SI Figure S4).
$XP-EHH$, which also makes use of haplotype information, has more
power than $F_{ST}$ but less than $hapFLK$ for hard sweeps detection. The
decrease in power for soft sweeps is also more pronounced for $XP-EHH$
(SI Figure S4), which is expected because $XP-EHH$ is designed to detect the
rise in frequency of one single haplotype.

Focusing on $hapFLK$, we further studied the evolution of the
detection power as a function of the initial and final frequencies of
the selected allele (SI Figure S5).  Although soft sweeps are
obviously harder to detect, there is still a reasonable power to detect
such events with $hapFLK$. For example, when the initial frequency is
20\% and the final frequency is 90\%, the detection power is greater
than 75\%, for a type I error rate of 1\%. When selection acts on
mutations at initial low frequency, the detection power is relatively
high (around 60\%) even for incomplete sweeps with a final frequency
of 50-60\%.

We also compared $FLK$, $hapFLK$ and $XP-EHH$ in simulations mimicking
data arising from full sequencing or imputation from a sequenced
reference panel.  This increase in marker density results in a
greater power for all tests (SI Figure S6). In this setting, $FLK$
is the most powerful. This comes from the fact that the selected
SNP, where the allele frequency difference between populations is
expected to be the largest, is always included in the sample in this
simulation setting.  In contrast, the selected SNP itself is often
missing when analyzing genotyping data, and information concerning
this SNP is then better captured by haplotypes than by single
neighboring SNPs. 
All results below were obtained on simulations mimicking dense
genotyping data.

\begin{figure}
\begin{center}
\includegraphics[width=3.27in]{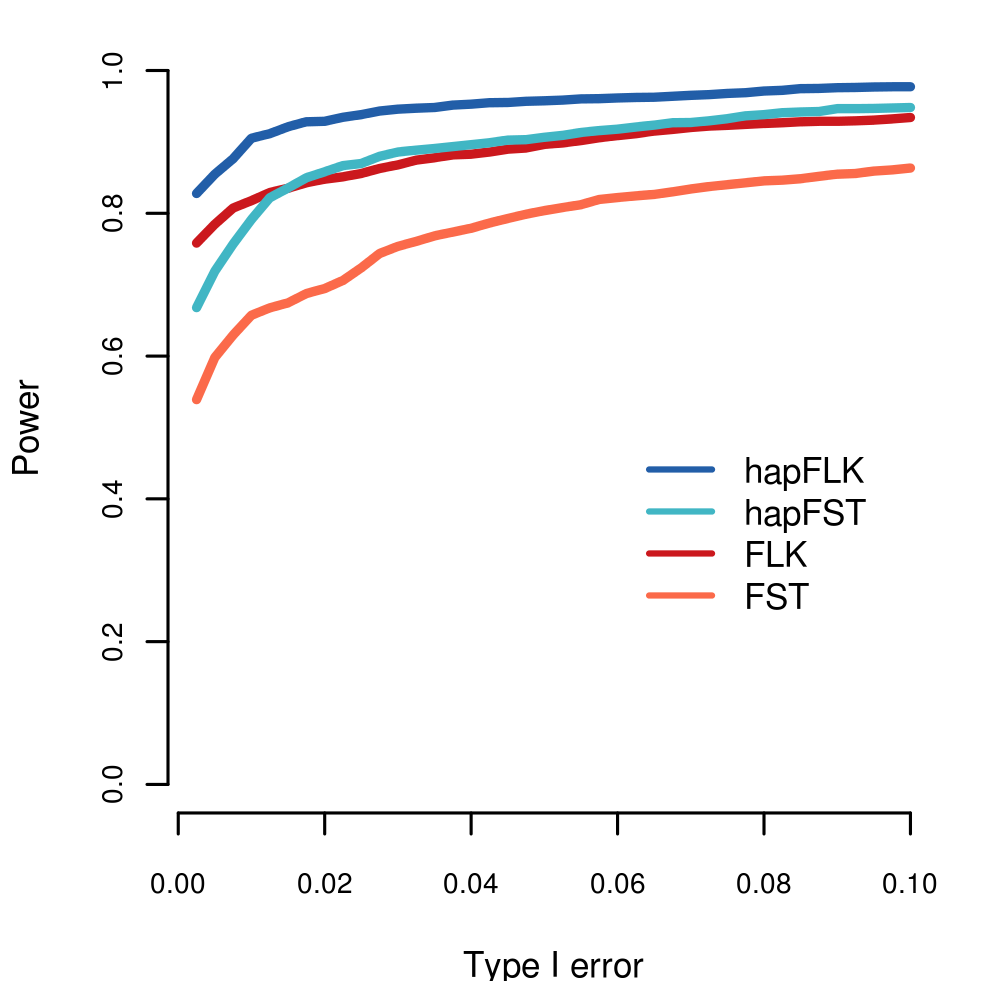}
\end{center}
\caption{
 \bf{Power of  $F_{ST}$, $FLK$, $hapF_{ST}$ and $hapFLK$ in the 4-population scenario as 
a function of the type I error rate. The initial frequency of the
selected allele is 1\%.}} 
\label{4pop}
\end{figure}

\paragraph{Hierarchical structure of populations:}
We then considered a four population sample, where populations are
hierarchically structured (Figure \ref{Figure1}, right).  This allows
to compare $hapFLK$ with related tests accounting for population
structure only ($FLK$), haplotype information only ($hapF_{ST}$), or
none of these features ($F_{ST}$).  As expected, the least powerful
approach in this scenario is the classical $F_{ST}$. The gain in power
provided by using a haplotype-based approach is of similar size as
that provided by accounting for population structure. Finally,
combining the two within a single statistic ($hapFLK$) results in an even greater
power gain (Figure \ref{4pop}).

A classical approach for selection scans based on more than two
populations is to perform tests on pairs of populations.  It is for
instance the only possible option for selection scans based on
$XP-EHH$.  To evaluate the interest of this pairwise strategy, we
compared the detection power obtained by applying $hapFLK$ on pairs of
populations or on the four populations jointly, and found that testing
all pairs of populations is always less powerful (SI Figure S7). Since
$XP-EHH$ also has less detection power than $hapFLK$ in the
two-population scenario, we can expect that applying $hapFLK$ using
the 4 populations jointly will be much more efficient than applying
$XP-EHH$ on pairs of populations.

\paragraph{Robustness and power of $hapFLK$ in complex demographic scenarios:}

The model underlying $hapFLK$ is that of pure drift evolution, with
constant population size in each branch of a population tree with no
admixture. These assumptions are made (i) when estimating the
population covariance matrix $\F$ and (ii) when assuming allele
frequency differences (either SNP or haplotype) are only due to
$\F$. We studied the robustness of $hapFLK$ in presence of admixture
or bottleneck events by examining separately their consequences on (i)
the estimation of the $\F$ matrix and (ii) the distribution of the
$hapFLK$ statistic.  For this, we simulated the evolution of two
populations with either a bottleneck in one of the population,
migration from one population to the other or migrations between both
populations (see Methods for details).

The estimation of the $\F$ matrix is slightly affected by
demographic events (SI Figure S2). When one of the population has
experienced a severe bottleneck (reduction in size of a factor 10),
the estimated branch length for this population is increased by
10\%. In the presence of migrations between populations, the two
branches remain of the same length but the Reynolds genetic distance
between the two populations is smaller than it should be (5\% smaller
in the one way migration case and 10\% smaller in the two-way
migration case).

Using this information we were able to perform simulations under pure
drift evolution or bottleneck / migration evolution that led to the
same \emph{estimated} $\F$ matrix. As $hapFLK$ is conditioned on this
estimate, this approach allows to evaluate the effect of demographic
events on the statistic, while integrating out their effect on $\F$.
We found that the distribution of $hapFLK$ was not greatly affected by
deviations from pure drift evolution, on par with $XP-EHH$ (SI Figure
S8).  Overall, these results show that while the estimate of $\F$ can
be affected by deviation from the evolution model, and therefore
coefficients in $\F$ must not be interpreted too literally, the
distribution of $hapFLK$ conditioned on this estimate is robust.
Besides, the power of $hapFLK$ is only slightly reduced under
migration scenarios and unchanged under a bottleneck scenario
(SI Figure S9).

\subsection*{Application to the Sheep Hapmap dataset} \label{DataApp}

To provide an insight into the advantages and issues of using $hapFLK$
on real data, we provide an example of application to a subset of the
data from the Sheep HapMap project. In sheep populations drift
accumulates rapidly, due to their small effective size, typically a
few hundred individuals \citep{kijas-etal-12}. As little power is
expected from analyses based on genetic differentiation if populations
are too distant, we focused on a group of relatively closely related
breeds from Northern European origin. Six populations are included in
this group, whose population tree is shown in Figure
\ref{fig:localtrees}, top left.

\begin{figure}
\begin{center}
\includegraphics[width=6.83in]{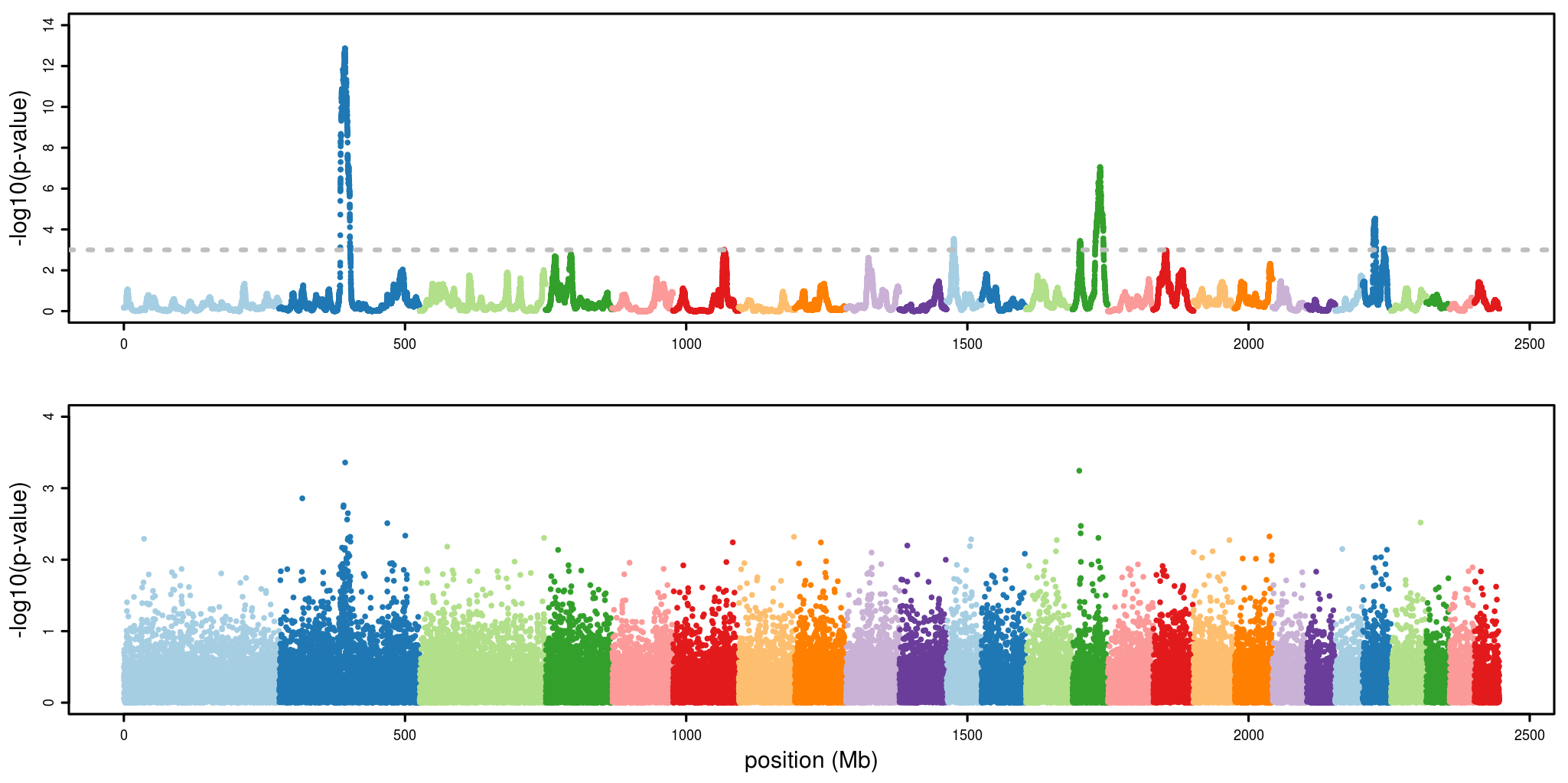}
\end{center}
\caption{ {\bf Genome scan for selection in Northern European Sheep using
  an haplotype-based ($hapFLK$, top) or single SNP ($FLK$, bottom)
  test.} x-axis : position on the genome, y-axis:-log10(p-value).}
\label{Figure6}
\end{figure}

\begin{figure}
  \centering
  \includegraphics[width=5in]{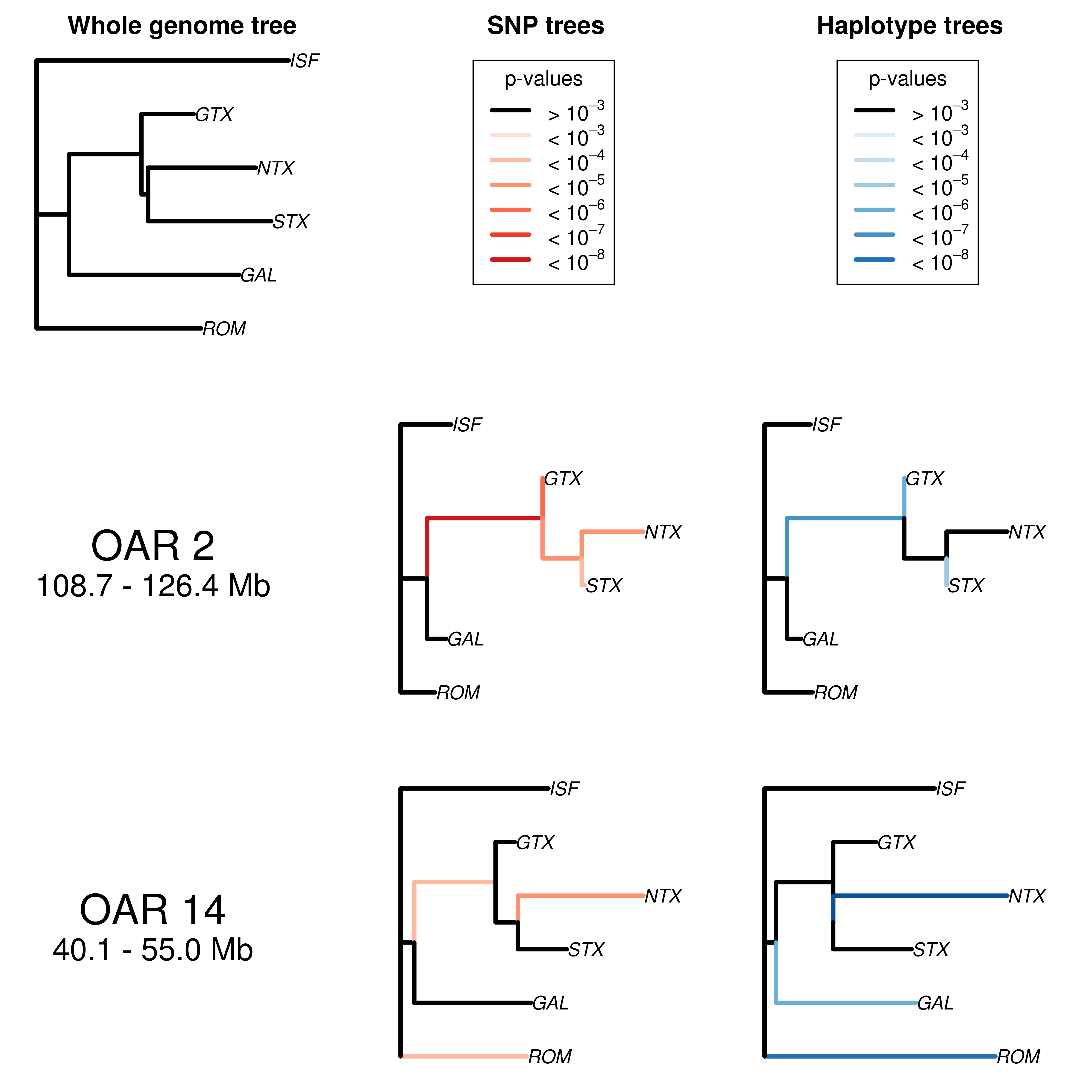}
  \caption{{\bf Local population trees estimated in two significant
      regions in the Sheep dataset.} Population tree of the Northern 
      European Sheep populations from the Sheep HapMap project (top left).
      Local population trees were estimated using Reynolds distance based on SNPs 
      (left) or haplotype clusters (right). Abbreviations: Irish Suffolk (ISF), German
      Texel (GTX), New Zealand Texel (NTX), Scottish Texel (STX), Galway (GAL), New
      Zealand Romney (ROM).
      }
  \label{fig:localtrees}
\end{figure}

The genome scan performed with $FLK$ provides little evidence for any
sweep in these data, with p-values of the order of $10^{-4}$, a hardly
convincing figure, only seen on chromosome 2 and 14. This is in great
contrast (Figure \ref{Figure6}) to the genome scan with $hapFLK$ which
identifies seven genome-wide significant regions (Table \ref{tab:EN}),
consistent with the additional power provided by $hapFLK$ on simulated
datasets.  For each of these regions, we identified the population(s)
under selection  by re-estimating the local
population trees and comparing it to the tree estimated from whole genome
data (see Methods for more details). 

 \begin{table}[ht]
  \caption{ {\bf Selective sweeps detected by $hapFLK$ within Sheep populations
    from Northern Europe}. For each significant region are listed: the chromosome region 
    (in Megabases on assembly OAR v2.0), the position of the maximum value for the statistic, 
    the corresponding p-value, the suspected
    selected population(s) along with selected haplotypes frequencies, and 
    potential candidate genes. {\small $\dagger$ the 
      signal in this region is not due to evolutive forces acting on the
      population, see details in Supporting Information.} }
{\small
\begin{tabular}{lllllll}
  region  & chr & position (Mb) & max (Mb) & p-value & population(s) (freqs) & candidate genes \\
  \hline
  1 & 2 & 108.7-126.3 & 116.9 & 1.5$\times 10^{-13}$ & STX (0.85), NTX (0.87), GTX (0.63) & GDF8 \\
  2 & 6 & 91.2-91.3 & 91.2 & 9.8$\times 10^{-4}$ & ROM (0.36, 0.32)  & \\ 
  3 & 11 & 12.6-14.0 & 13.7 &  4.2$\times 10^{-4}$ & ROM (0.75), GAL (0.45) & \\ 
  4 & 14 & 12.2-14.6 & 13.9 & 4.5$\times 10^{-4}$ & ISF (0.65) & \\ 
  5 & 14 & 40.1-55.0 & 48.8 & 8.8$\times 10^{-8}$ & ROM (0.29, 0.44), NTX (0.54) & TFGB1, IRF3 \\ 
  6 & 22$\dagger$ & 19.1-24.0 & 21.7 & 5.5$\times 10^{-5}$ & GTX (0.62) & PITX3 \\ 
  7 & 22 & 38.5-38.8 & 38.6 & 8.6$\times 10^{-4}$ & ROM (0.31, 0.35) & \\ 
\end{tabular}
}
\label{tab:EN}
\end{table}

Figure \ref{fig:localtrees} shows local trees for the two largest
signal, on chromosome 2 and 14 (local trees for the other significant
regions are provided in SI Figure S10).

\begin{figure}
\begin{center}
\includegraphics[width=6.83in]{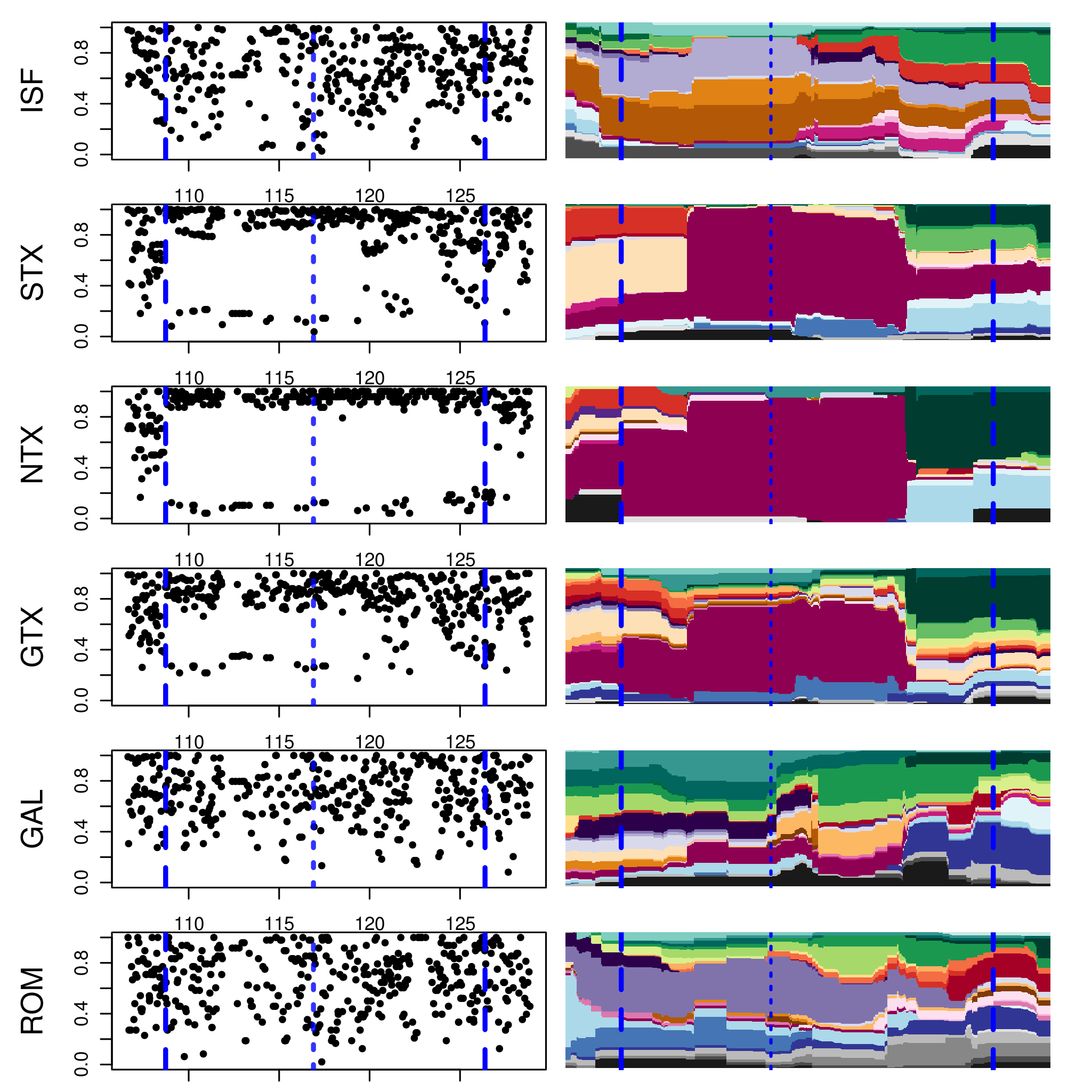}
\end{center}
\caption{ {\bf Allele (left) and haplotype cluster (right) frequencies
    in detected region 1 (Chromosome 2) for each of the 6 Sheep
    populations used in the test.} Blue bars indicate the limits of
  the detected region and the position of maximum of the test.}
\label{fig:freqGDF8}
\end{figure}

The most significant selection signature (region 1 in Table 1)
corresponds to a 15 Mb region in chromosome 2. Selection occurred in
the three Texel breeds, most likely acting on the myostatin gene
\emph{GDF-8}, which is located in the middle of the region. Indeed,
Texel sheep carries a mutation in this gene, which contributes to
muscle hypertrophy \citep{GDF8}, a strongly selected trait in these
populations.  Although the mutation was discovered in Belgian Texels,
our results imply that it must be present in these other Texel
populations.  The $F_{ST}$ genome scan performed by
\citet{kijas-etal-12}, which was based on single SNP tests, already
detected a selection signature in region 1. SNPs within this region
are almost fixed in the three Texel populations (Figure
\ref{fig:freqGDF8}), evidencing a hard sweep signal. However, even in
this ``easy'' case, using haplotype information makes the detection
signal more interpretable: while $FLK$ only exhibits moderate p-value
decrease in the region, from which no clear conclusion concerning the
selected site position can be drawn, $hapFLK$ provides a continuous
and strong signal covering the whole region and almost centered on the
selected site. The local tree exhibits a large increase in branch
length in the branch ancestral to the three Texel populations, and
reduced branch length between Texel populations (Figure
\ref{fig:localtrees}). This is consistent with a shared selection
event predating the split between populations. Finally, the example of
region 1 also illustrates that our test can detect selection
signatures that are shared by several populations, which we did not
formally test in the simulations.
 
In contrast to the selection signature around \emph{GDF-8}, the second
most significant region (region 5, on chromosome 14) shows no evidence
of a hard sweep (Figure \ref{Figure6}) and cannot be identified using
the single marker $FLK$ test. The local tree (Figure
\ref{fig:localtrees}) computed using SNP data exhibits slightly
increased branch lengths, whereas the local tree computed using
haplotype clusters presents very strong evidence for selection in two
breeds: the New Zealand Texel and the New Zealand Rommey, together
with reduced haplotype diversity (SI Figure S14).  These two breeds
are not historically closely related (Figure \ref{fig:localtrees}, top
left), but both have been imported in New Zealand (in 1843 and 1991
respectively). The selection signature could thus be due to a common
recent selection pressure on the two breeds in the last decades. This
would be consistent with the relatively modest frequency of the
selected clusters, and the fact that these selected clusters are
different in the two breeds, suggesting that selection started on
different haplotype backgrounds. One possible underlying trait
associated with this selection signal is resistance to nematode-like
parasites, an important disease affecting sheep in New Zealand. Two
studies \citep{FECQTL,TGFB1} found evidence for association between
genetic polymorphism and parasite resistance related traits in this
region of the genome in Texel breeds. \citet{FECQTL} also found these
polymorphisms associated with muscle depth. While the functional basis
of these two effects is still unclear (pleiotropy, linkage
disequilibrium with growth factors), it is possible that animal
fitness in this region is related to multi-locus haplotypes rather
than to single SNPs.

We point the reader interested in details for all significant regions
in Table 1 to the Supporting Information. In particular, allele and
haplotype cluster frequencies are provided in Figures S11-S16 and
local trees in Figure S10. An alternative approach for pinpointing the
selected population(s) is also described (section 1.2) and applied to
these regions (section 2.3, Figures S17-S21).


\section*{DISCUSSION}

\paragraph{Haplotype versus single marker differentiation tests:} 
For the analysis of dense genotyping data, where the selected site
itself is generally not observed, we showed that using haplotypes
rather than single SNPs greatly improves the
detection power of selection signatures.  This was not the case for
the analysis of sequencing data.  However, our simulations involved
one single selected site, which is the most favorable situation for
single SNP approaches as $FLK$.  In many real situations, selection
will rather act effectively on multi-locus haplotypes
\citep{PRIPIC2010}, due for instance to recurrent mutations affecting
the same gene, or to polygenic selection.  We expect haplotype based
tests to be more powerful in such situations, which according to us
justifies their use also for the analysis of sequencing data.  In the
particular case of low coverage re-sequencing, which is becoming a
common experimental design in population genetics, this analysis will
have to account for the additional uncertainity in genotype
estimation, but we believe this can easily be tackled by the
clustering algorithm used for $hapFLK$.

\paragraph{Different strategies for the inclusion of haplotype information in differentiation:} 
To extend the single marker $FLK$ and $F_{ST}$ tests into haplotype
based tests, we decided to estimate local haplotype clusters from
genotype data and to consider these estimated clusters as alleles.  An
alternative strategy could be to construct these haplotype based
statistics by choosing a genomic window and computing haplotype
frequencies in this window for each population. But this direct
approach has several drawbacks. First, haplotypes are generally
unknown and have to be inferred from genotypes, which typically relies
on a model for Linkage Disequilibrium (LD) such as that of
\citet{FPh}. Using directly the model parameters as we do has the
advantage of allowing to average over the uncertainty in the
distribution of possible haplotypes rather than using a best guess
which is known to include errors \citep{marchini.etal.2006}. Second,
for the direct approach suitable values of window size and window
overlap have to be found. These values will likely depend on the
patterns of LD along the genome, which are known to vary. Using a
Hidden Markov Model for LD as we do eliminates the need to define such
windows and naturally incorporates variation in LD along the genome.
Finally, several similar haplotypes may be associated to the same
selected allele, and treating them independently should affect the
detection power of the tests. In the fastPHASE model, similar
haplotypes are clustered together and will be considered as a single
allele.  

In our implementation of $hapFLK$, we decided to fit
haplotype clusters using the fastPHASE algorithm \citep{FPh}. Other
haplotype clustering models, for instance Beagle \citep{BEAGLE}, could
certainly be used as well.  For example, the pattern of haplotype
frequencies around the {\it LCT} gene in human populations was studied
using either fastPHASE \citep{JAC2008} or Beagle \citep{Browning}, and
a strong evidence for selection in Europe was observed in both cases.
However, to go beyond these observations and build a formal
statistical test for selection, it is important to realize that the
distribution of $hapFLK$ (or $hapF_{ST}$) depends on the number of
clusters used to model haplotype diversity.  This number is fixed in
fastPHASE but variable along the genome in Beagle.  As this variation
might be due to natural selection, but also to other effects such as
variations in recombination or mutation rate, further studies would be
required to evaluate the influence of using different clustering
algorithms on the detection power.

Another important feature of $hapFLK$ is its ability to account for
the hierarchical structure of the sampled populations, arising from
their evolutionary history within the species. $FLK$ was already shown
to be more powerful than the $F_{ST}$ test in many simulated scenarios
\citep{Maxime}. It was also compared to a popular Bayesian
differentiation test \citep{FollGag} in one simulated scenario with
hierarchically structured populations, and again provided more
detection power. Consequently, we expect that $hapFLK$ will also
perform better than other haplotype based differentiation tests
\citep{Gompert,Guo,Browning} for hierarchically structured
populations.

In order to build tests that account for both the differentiation
between populations and haplotype structure, all methods discussed
above propose to include haplotype information into single marker
differentiation tests.  Another popular strategy, developed in the
XP-EHH \citep{XP-EHH} and Rsb \citep{Rsbi} statistics, is to compute a
statistic quantifying the excess of long haplotypes within each
population, and to contrast this statistic among pairs of populations.
Simulating a two population sample, we found that XP-EHH and $hapFLK$
had relatively similar power for hard sweep detection.  However, one
important difference was that $hapFLK$ maintained some power for soft
sweep detection, in contrast to XP-EHH.

When more than two populations are sampled, comparing only pairs of
populations raises a multiple testing issue leading to a significant
decrease in power (SI Figure S7).  Besides, computing a single test
at the meta population level seems more appropriate for several
reasons.  First, the signals we detected in sheep suggest that
favorable alleles are often positively selected in several
populations, either closely (region 1) or distantly related (region
5).  Second, our ability to detect loci under selection depends on our
ability to estimate the allele frequencies in this common ancestral
population, which is clearly improved when using all populations
simultaneously.  One potential difficulty arising from our
meta-population approach is the identification of the population(s)
under selection, which is more difficult than when comparing pairs of
populations. We proposed to address this question using a local
re-estimation of the population tree, as illustrated in the Sheep
Hapmap data analysis.  An alternative approach, which is based on a
spectral decomposition of $hapFLK$, is also described in the
Supporting Information and applied to the Sheep data.

\paragraph{Robustness of $hapFLK$ and computation of p-values in a general situation}
In many genome scans for selection, all loci above a given empirical
quantile of the test statistic are considered as potential targets of
selection.  However, this so-called outlier approach does not allow to
control the false positive rate and can be inefficient in many
situations \citep{Teshima}.  To overcome this limitation and quantify
the statistical significance of selection signatures, one needs to
describe the expected distribution of the test statistic under neutral
evolution, which depends on the demographic history of the sampled
populations.  In the case of $hapFLK$, this neutral distribution could
be estimated by (i) fitting the kinship matrix $\F$ from genome wide
SNP data and (ii) simulating neutral samples conditional on $\F$,
using a simple model with no migration and constant population size
along each branch of the population tree.  This approach avoids to
estimate a full demographic model for the sampled populations, and was
found to be robust to bottlenecks or to intermediate levels of
migration / admixture (SI Figure S8).  For the analysis of samples
involving stronger departures from the hierarchical population model
assumed in this study (for instance with hybrid populations), the
expected covariance matrix of allele frequencies could also be
modelled using relaxed hypotheses.  The strategies used in Bayenv
\citep{Coop} or TreeMix \citep{PIC2012} could for instance be adapted
to the application of $hapFLK$.

However, in many situations (\emph{e.g.} in the Sheep HapMap data),
the neutral distribution of $hapFLK$ is not only affected by
demography, but also by SNP ascertainment bias.  Simulating the
ascertainment process is in general difficult, in the Sheep data for
example it involves animals from a large panel of worldwide
populations \citep{kijas-etal-12}.
For single SNP tests such as $FLK$, this ascertainment issue can be
circumvented by estimating a neutral distribution for several bins of
the allele frequency in the ancestral population \citep{Maxime},
because we can assume that the only effect of SNP ascertainment is to
bias the allele frequency distribution. But this strategy is not
applicable to haplotype based tests, for which the effect of SNP
ascertainment is more complex.  We consequently proposed a more
empirical approach, where the null distribution of $hapFLK$ is
directly estimated from the data using an estimator that is robust to
outlier values.  This empirical approach might be useful in future
genome scans for selection, even if they are based on different test
statistics than $hapFLK$, but its validity will depend on each
particular dataset and needs to be checked carefully by looking at the
p-value distribution (see Methods for more details).

The most significant selection signatures detected in Sheep using
$hapFLK$ exhibit extremely small p-values (down to $10^{-13}$), while
the smallest p-values obtained with $FLK$ for the same dataset were of
order $10^{-4}$. This difference of magnitude might be artificially
inflated by the fact that we compute $hapFLK$ p-values using a normal
distribution, and $FLK$ p-values using a chi-square distribution.
However, we note that the choice of these distributions is supported
by the data.  Besides, we found that $FLK$ p-values in simulated
samples with selection using a chi-square distribution can go down to
at least $10^{-11}$ (data not shown).  We thus believe that the
p-value difference observed in sheep reflects the fact that $hapFLK$
is indeed much more powerful than $FLK$, especially for SNP data where
ascertainment bias leads to remove SNPs with extreme allele
frequencies.

\paragraph{Soft or incomplete sweeps:} While genome scans for
selection have historically focused on hard sweeps, several recent
studies have pointed out the importance of soft sweeps in the
evolution of populations \citep{Przeworski, PRIPIC2010} and described
the genomic signature of these selection scenarios \citep{Hermisson}.
We tested $hapFLK$ for initial frequencies of the favorable allele up
to 30\%, and found that reasonable power could be achieved also in
this situation.  The detection of incomplete sweeps is another
important issue, which has not been much tackled in the literature.
Indeed, detecting selected alleles at intermediate frequency is almost
impossible with methods based on the allele frequency spectrum, and
very difficult with EHH or $F_{ST}$ based existing approaches.  In
contrast, $hapFLK$ is quite powerful in the case of incomplete
sweeps, and several of the selection signatures detected in the Sheep
HapMap data correspond to intermediate frequencies of the selected
haplotype (see Table 1).

Few hard sweeps were actually detected in the Sheep data, although
they are easier to detect than soft sweeps.  This might be due to the
short divergence time between these populations (a few hundred
generations), which would limit the rise in frequency of favorable
alleles. On the other hand, artificial selection has been associated
with strong selection intensities, especially in the last decades,
which should compensate for the short evolution time.  One alternative
explanation could be the variation of the selection intensity or
direction over time, due to changes in agronomical objectives
(\emph{e.g.} in the Sheep from wool to meat production) or
importations of animals in a new environment (\emph{e.g.} in the Texel
and Romney breeds from Europe to New Zealand).  The small number of
hard sweeps can also be explained by the fact that artificial
selection on quantitative traits is in general polygenic.//

As a final and general remark on all methods aiming at discovering positive selection,  
selective constraints in functional and
non-functional regions are probably more complex than what is usually simulated (with purifying
and background selection, polygenic selection, balancing selection, etc).
Definitely more research effort needs to be done on these aspects.

\paragraph{Conclusions:} Overall, our study demonstrates that using
haplotype information in $F_{ST}$ based tests for selection greatly
increases their detection power.  Consistent with several recent other
studies \citep{Maxime,Excoffier,Coop}, it also confirms the importance
of analyzing multiple populations jointly, while accounting for the
hierarchical structure of these populations. The new $hapFLK$
statistic, which combines these two features, can detect a wide range
of selection events, including soft sweeps, incomplete sweeps, sweeps
occuring in several populations, and selection acting directly on
haplotypes. 


\section*{Acknowledgments}

The ovine SNP50 HapMap dataset used for the analyses described was
provided by the International Sheep Genomics Consortium and obtained
from www.sheephapmap.org in agreement with the ISGC Terms of Access.
The simulations and data analysis were performed on the computer
cluster of the bioinformatics platform Toulouse Midi-Pyrenees.  We
thank Michael Blum and Lucia Spangenberg for useful comments on the
manuscript and Carole Moreno for earlier access to the Sheep HapMap
data.

\clearpage

\bibliography{hapflk}

\end{document}